\documentclass[manuscript,twocolumn]{aastex}
\usepackage{spr-astr-addons}
\begin{document}

%\setpagewiselinenumbers \modulolinenumbers[1] \linenumbers

\title{Relativistic compact stars in $f(T)$ gravity admitting conformal motion}
\shorttitle{Relativistic compact stars} \shortauthors{Das et al.}

\author{Amit Das\altaffilmark{1}}
\altaffiltext{1}{Department of Physics, Indian Institute of
Engineering Science \& Technology, Shibpur, Howrah 711103, West
Bengal, India\\ amdphy@gmail.com}

\author{Farook Rahaman\altaffilmark{2}}
\altaffiltext{2}{Department of Mathematics, Jadavpur University,
Kolkata 700 032, West Bengal, India\\ rahaman@iucaa.ernet.in}

\author{B.K. Guha\altaffilmark{3}}
\altaffiltext{3}{Department of Physics, Indian Institute of
Engineering Science \& Technology, Shibpur, Howrah 711103, West
Bengal, India\\ bkguhaphys@gmail.com}

\author{Saibal Ray\altaffilmark{4}}
\altaffiltext{4}{Department of Physics, Government College of
Engineering \& Ceramic Technology, Kolkata 700 010, West Bengal,
India\\ saibal@iucaa.ernet.in}

\begin{abstract}
We provide a set of exact spherically symmetric solutions
describing the interior of a relativistic star under $f(T)$
modified gravity. To tackle the problem with lucidity we also
assume the existence of a conformal Killing vector field within
this $f(T)$ gravity. We study several cases of interest to explore
physically valid features of the solutions.
\end{abstract}


\section{Introduction}
To search and understand deeply physical aspects of the
astrophysical and cosmological phenomena there still remain
several challenging and intrigued problems for the theoretical
physicists. Einstein's theory of gravitation, though has been
always a very fruitful tool for uncovering hidden mysteries of
Nature, does not meet all the criteria to explain the current
paradigm of astrophysics and cosmology. As a result, several
alternative theories of gravity have been proposed time to time.
Recently, two generalized functional theories, namely, $f(R)$
gravity and $f(T)$ gravity, received more attention as an
alternative theories of Einsteinian gravitational theory.

However, $f(T)$ theory of gravity is more controllable than $f(R)$
theory because the field equations in the former one are
restricted to the second order differential equations whereas the
later case is relatively difficult to handle as those are of
fourth order differential equations \citep{Boehmer2011}. Obviously
one can take help from the Palatini approach to make $f(R)$ theory
a second order system of differential equations
\citep{Durrer2010,Felice2010,Sotiriou2010}. Another general
feature of $f(T)$ gravity is that its construction is associated
with a generalized Lagrangian \citep{Bengochea2009,Linder2010}.

In connection to application of $f(T)$ gravity we note that most
research have been oriented to cosmology - theoretical
presentation as well as observational
verification~\citep{Wu2010a,Tsyba2011,Dent2011,Chen2011,Bengochea2011,Wu2010b,Yang2011,Zhang2011,Li2011,Wu2011,Bamba2011}.
However, later on astrophysical applications also can be observed
in the following
works~\citep{Boehmer2011,Deliduman2011,Wang2011,Daouda2011,Abbas2015}.
Among these works we are specially interested to the works
of~\citep{Boehmer2011,Deliduman2011}. \citet{Deliduman2011}
claimed that relativistic stars in $f(T)$ theory of gravity do not
exist whereas \cite{Boehmer2011} proved that they do exist. In the
present work we shall follow the result of \citet{Boehmer2011} by
considering the problem in $f(T)$ gravity along with conformal
Killing vector field. There are some other standard works on
$f(T)$ gravity which one may consult for further studies with
various physical aspects
~\citep{Andrade2000,Ferraro2011,Tamanini2012a,Tamanini2012b,Aldrovandi2013,Aftergood2014}.

The conformal Killing vectors (CKV) provide inheritance symmetry
which conveniently makes a natural relationship between geometry
and matter through the Einstein field equations. In favor of the
prescription of this mathematical technique CKV,
\cite{Rahaman2015a} mention the following features: (1) it
provides a deeper insight into the spacetime geometry and
facilitates the generation of exact solutions to the Einstein
field equations in a more comprehensive forms, (2) the study of
this particular symmetry in spacetime is physically very important
as it plays a crucial role of discovering conservation laws and to
devise spacetime classification schemes, and (3) because of the
highly non-linearity of the Einstein field equations one can
reduce easily the partial differential equations to ordinary
differential equations by using CKV.

Therefore, several works have been performed by use of the
technique of conformal motion as can be seen available in the
literature. A few interesting applications of conformal motion to
astrophysical field can be found in the following Refs.
\citet{Ray2008},
\citet{Rahaman2010a,Rahaman2010b,Rahaman2014,Rahaman2015b,Rahaman2015c},
\citet{Usmani2011} and \citet{Bhar2014}. In this line of
investigations some special mention are of the very recent works
of \citet{Rahaman2015a} and \citet{Bhar2015}. Search for a new
wormhole solution inspired by noncommutative geometry with the
additional condition of allowing conformal Killing vectors was
performed by \citet{Rahaman2015a} whereas \citet{Bhar2015} have
shown that a new class of interior solutions for anisotropic stars
are possible by admitting conformal motion in higher-dimensional
noncommutative spacetime. Interior solutions admitting conformal
motions also have been studied extensively in the past by Herrera
and his co-workers
\citep{Herrera1984,Herrera1985a,Herrera1985b,Herrera1985c}.

Under this background, therefore, in the present work
our main aim or motivation is to construct
a set of stellar solutions under $f(T)$ theory of gravity
by admitting conformal motion of Killing Vectors. The
outline of the investigation is as follows: in the Sec. 2 we
provide the basic mathematical formalism of $f(T)$ theory whereas
in Sec. 3 the CKVs are formulated. The Einstein field equations
under $f(T)$ gravity and CKVs have been provided along with
their solutions in Sec. 4. We have tested physical validity
of the model by using Tolman-Oppenheimer-Volkoff (TOV) equation
in Sec. 5. Lastly, in Sec. 6 we pass some concluding remarks.

\section{The $f(T)$ theory: Basic Mathematical Formalism}

The action of $f(T)$ theory is taken as (for geometrical units
G=c=1)
\begin{equation}
S[e^i_\mu , \phi_A ] = \int d^4x \left[ \frac{1}{16 \pi} f(T) +
L_{matter} (\phi_A)  \right]\label{eq1}.
\end{equation}

Here, $\phi_A$ indicates matter fields and $f(T)$ is an arbitrary
analytic function of the torsion scalar $T$. The torsion scalar is
constructed from torsion and cotorsion as follows:
\begin{equation}
T = S_\sigma^{ \mu\nu} T^\sigma_{ \mu\nu},\label{eq2}
\end{equation}
where
\begin{equation}
T^\sigma_{ \mu\nu} =   \Gamma^\sigma_{ \mu\nu}-\Gamma^\sigma_{
\nu\mu} = e_i^\sigma \left( \partial_\mu e^i_\nu - \partial_\nu
e^i_\mu \right),\label{eq3}
\end{equation}

\begin{equation}
K_\sigma^{ \mu\nu} =  -\frac{1}{2} \left( T^{ \mu\nu}_{~~~\sigma}-
T_{~~~\sigma}^{ \nu\mu}-T_{\sigma}^{ \mu\nu}  \right),\label{eq4}
\end{equation}
are torsion and cotorsion respectively with newly defined tensor
components
\begin{equation}
S_\sigma^{ \mu\nu} =  \frac{1}{2} \left( K^{ \mu\nu}_{~~~\sigma}+
\delta_\sigma^\mu T ^{ \beta \nu}_{~~~\beta}-\delta_\sigma^\nu T
^{ \beta \mu }_{~~~\beta} \right).\label{eq5}
\end{equation}

Here, $e^i_\mu$ are the tetrad by which it is possible to define
any metric as $g_{\mu \nu} =  \eta_{ij} e^i_\mu e^j_\nu$, where $
\eta_{ij} = diag ( -1,1,1,1 )$ and $e_i^\mu e^i_\nu =
\delta_\nu^\mu $, $ e= \sqrt{-g} = det(e^i_\mu)$.

If one varies the action (\ref{eq1}) with respect to the tetrad,
one can get the field equations of $f(T)$ gravity which can be given by
\begin{equation}
S_i^{ \mu \nu}f_{TT} \partial_\mu T + e^{-1} \partial_\mu ( e
S_i^{ \mu \nu}) f_T -T^\sigma_{ \mu i}S_\sigma^{ \nu \mu} f_T +
\frac{1}{4} e_i^\nu f  = 4 \pi \Upsilon_i^\nu,\label{eq6}
\end{equation}
where
\[ S_i^{ \mu \nu} =e_i^\sigma S_\sigma^{ \mu \nu}, ~~f_T =
\frac{\partial f}{\partial T}~~~\&  ~~~f_{TT} = \frac{\partial^2
f}{\partial T^2}~~\] and $\Upsilon_i^\nu$ denotes the energy
stress tensor of the anisotropic fluid as
\[\Upsilon_\nu^\mu=  ( \rho + p_t)u^{\mu}u_{\nu} - p_t g^{\mu}_{\nu}+
            (p_r -p_t )\eta^{\mu}\eta_{\nu},\]
with $$ u^{\mu}u_{\mu} = - \eta^{\mu}\eta_{\mu} = 1. $$

\section{Conformal Killing Vector}

To search for a natural relationship between geometry and matter
Einstein's general relativity provides a rich arena to use
symmetries. Various symmetries that arising either from
geometrical point of view or physical relevant quantities are
known as collineations. The greatest advantageous collineations is
the conformal Killing vectors (CKV) which provide a deeper insight
into the spacetime geometry. In mathematical point of view,
conformal motions or conformal Killing vectors (CKV) are motions
along which the metric tensor of a spacetime remains invariant up
to a scale factor. Another advantage to use the CKV is that it
facilitates generation of exact solutions to the field equations.
This is achieved by reducing the highly nonlinear partial
differential equations of Einstein's gravity to ordinary
differential equations through the technique of CKV.

The CKV can be defined as
\begin{equation}
L_{\xi} g_{ij} = \xi_{i;j}+ \xi_{j;i} = \psi g_{ij},\label{eq7}
\end{equation}
where $L$ is the Lie derivative operator of the metric tensor and
$\psi$ is the conformal factor. It is supposed that the vector
$\xi$ generates the conformal symmetry and the metric $g$ is
conformally mapped onto itself along $\xi$. However, it is to note
that neither $\xi$ nor $\psi$ need to be static even though one
considers a static metric~\citep{Harko1,Harko2}. Further, one
should note that (i) if $\psi=0$ then Eq. (7) gives the Killing
vector, (ii) if $\psi=$ constant it gives homothetic vector, and
(iii) if $\psi=\psi(\textbf{x},t)$ then it yields conformal
vectors. Moreover, for $\psi=0$ the underlying spacetime becomes
asymptotically flat which further implies that the Weyl tensor
will also vanish. Therefore, CKV has an intrinsic property to
provide a deeper insight of the underlying spacetime geometry.
Basically the Lie derivative operator $L$ describes the interior
gravitational field of a stellar configuration with respect to the
vector field $\xi$.

Let us therefore assume that our static spherically symmetric
spacetime admits an one parameter group of conformal motion, so
that the metric
\begin{equation}
ds^{2}=-e^{\nu(r)}dt^{2}+e^{\lambda(r)}dr^{2}+r^{2}(d\theta^{2}+sin^{2}\theta
d\phi^{2}),\label{eq8}
\end{equation}
is conformally mapped onto itself along $\xi$.

Here Eq. (\ref{eq7}) implies that
\begin{equation}
L_\xi g_{ik} =\xi_{i;k}+ \xi_{k;i} = \psi g_{ik},\label{eq9}
\end{equation}
with $\xi_i = g_{ik}\xi^k$.

From Eqs. (\ref{eq8}) and (\ref{eq9}), one can get the following
expressions
\citep{Herrera1985a,Herrera1985b,Herrera1985c,Boehmer2011}
\begin{eqnarray}
&\xi^1 \nu^\prime =\psi,\nonumber \\
&\xi^4  = {\rm constant},\nonumber \\
&\xi^1  = \frac{\psi r}{2},\nonumber \\
&\xi^1 \lambda ^\prime + 2 \xi^1 _{,1} =\psi, \nonumber
\end{eqnarray}
where $1$ and $4$ stand for the spatial and temporal coordinates
$r$ and $t$ respectively.

The above set of equations imply
\begin{eqnarray}
e^\nu  &=& C_0^2 r^2, \label{eq10}\\ e^\lambda  &=& \left[\frac
{C} {\psi}\right]^2,  \label{eq11} \\ \xi^i &=& C_1 \delta_4^i +
\left[\frac{\psi r}{2}\right]\delta_1^i, \label{eq12}
\end{eqnarray}
where $C$, $C_0$ and $C_1$ all are integration constants.

\section{The Field equations and their solutions}
Let us define tetrad matrix for the metric (\ref{eq8}) as follows:
\begin{equation}
e^i_{ \mu} = ~diag [ \sqrt{C_0^2 r^2}, e^{\frac{\lambda}{2}}, r, r
\sin\theta ].\label{eq13}
\end{equation}

Therefore, the torsion scalar can be determined as
\begin{equation}
T(r) = - \frac{6 e^{-\lambda}}{r^2}.\label{eq14}
\end{equation}

Inserting this and the components of the tensors $ S_i^{ \mu \nu}$
and $T_i^{ \mu \nu}$ in Eq. (\ref{eq6}) we obtain
\begin{equation}
4 \pi \rho = \frac{ e^{-\lambda}}{r} T^{\prime} f_{TT} - \left [
\frac{T}{2} + \frac{ 1}{2r^2} + \frac{ e^{-\lambda}}{2r} \left(
\frac{2}{r} + \lambda^\prime \right) \right]f_T +
\frac{f}{4},\label{eq15}
\end{equation}

\begin{equation}
4 \pi p_r = \left [ \frac{T}{2} + \frac{ 1}{2r^2} \right]f_T -
\frac{f}{4},\label{eq16}
\end{equation}

\begin{equation}4 \pi p_t =  -\frac{e^{-\lambda}}{r} T^{\prime} f_{TT} + \frac{T }{4} f_T
  -\frac{e^{-\lambda}}{2} \left [\frac{1}{ r^2}
  -\frac{\lambda^\prime }{r} \right]f_T -\frac{f}{4}.\label{eq17}
\end{equation}

Due to \cite{Boehmer2011}, one of the possible way to get back
general relativistic result is $f_{TT} = 0$, although its general
relativity form has no meaning in the present context. Therefore,
we are now seeking solutions using
\begin{equation}
f(T) = T \label{eq18a}
\end{equation}
and
\begin{equation}
f(T)=aT+b,\label{eq18b}
\end{equation}
which immediately follows from $f_{TT} = 0$. Here $a$ and $b$ are
two purely constant quantities. For a fluid distribution
consisting of normal matter, we have
\begin{equation}
p_r = \omega \rho,\label{eq19}
\end{equation}
where $\omega$ is an equation of state parameter.

\subsection{CASE I: $f(T)=T$}

\subsubsection{$p_r = p_t = p$}

Now, using Eqs. (\ref{eq14}) - (\ref{eq18a}), we obtain the
following solutions \citep{Herrera1984,Herrera1985c}:
\begin{equation}
e^{-\lambda(r)}=\left( \frac{1}{2}+ F r^2\right),\label{eq20}
\end{equation}

\begin{equation}
T(r) = - \left[\frac{3}{r^2}+6F\right],\label{eq21}
\end{equation}

\begin{equation}
4 \pi \rho(r)= -\frac{1}{4r^2}+\frac{3}{2}F,\label{eq22}
\end{equation}

\begin{equation}
4 \pi p =-\frac{1}{4r^2}-\frac{3}{2}F ,\label{eq23}
\end{equation}

\begin{equation}
\psi=  C \left[ \frac{1}{2}+Fr^2\right]^{\frac{1}{2}},\label{eq24}
\end{equation}
where $F$ is an integration constant.

\begin{figure}[h]
\centering
\includegraphics[width=0.3\textwidth]{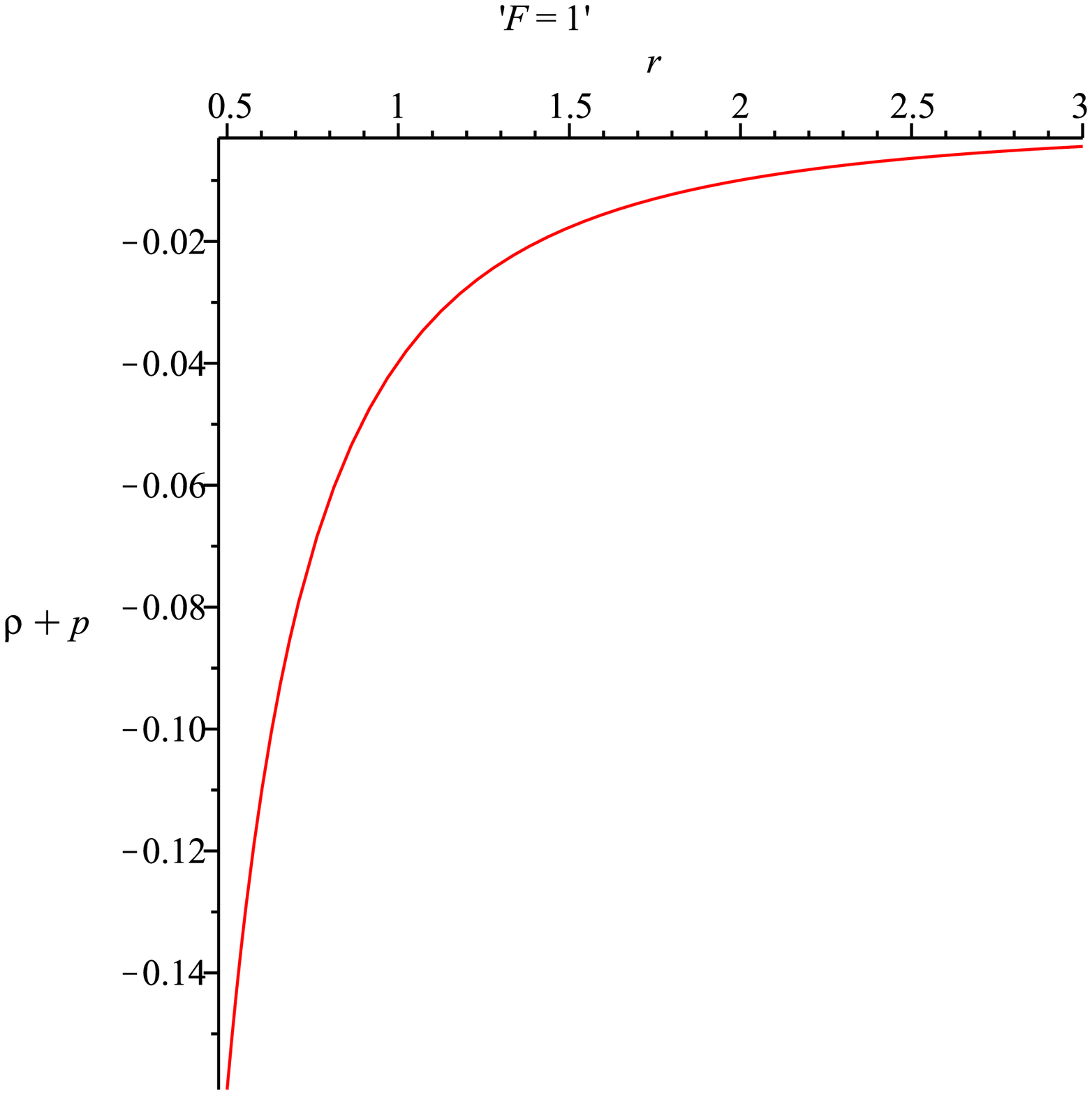}
\includegraphics[width=0.3\textwidth]{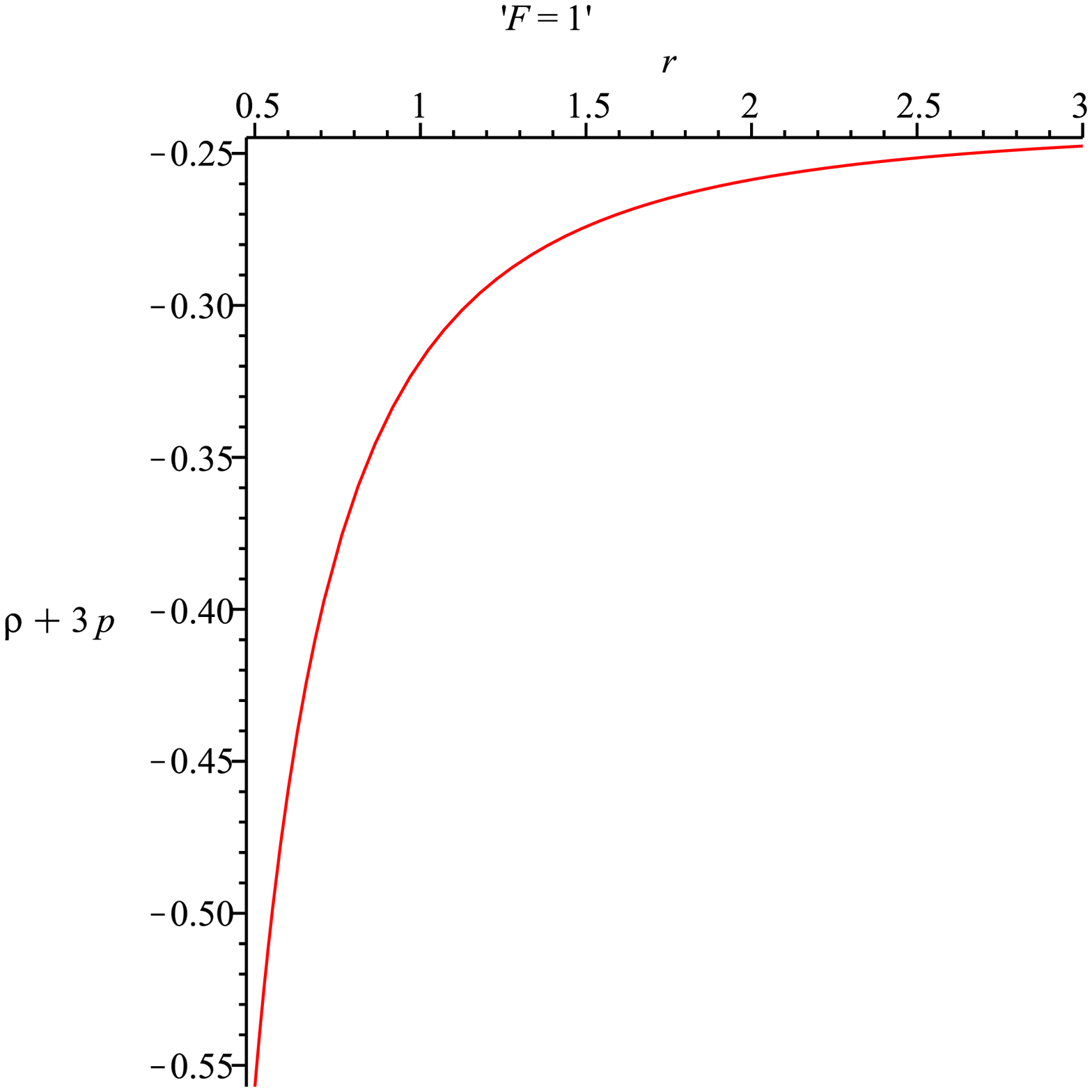}
\caption{Variation of $\rho + p$ (Top) and $\rho + 3p$ (Bottom)
are shown with respect to radial coordinate for isotropic case}
\end{figure}

The sound velocity ${v_s}^2$ can be found out as
\begin{equation}
\frac{dp}{d\rho}= 1.\label{eq25}
\end{equation}

\subsubsection{$p_r \neq p_t$}

Now, using Eqs. (\ref{eq14}) - (\ref{eq18a}) and (\ref{eq19}), we
obtain the following solutions \citep{Herrera1984,Herrera1985c}:
\begin{equation}
e^{-\lambda(r)}=  \left[\frac{\omega+1}{\omega+3}\right]
+Dr^{-\left[\frac{\omega+3}{\omega}\right]},\label{eq26}
\end{equation}

\begin{figure}[h]
\centering
\includegraphics[width=0.3\textwidth]{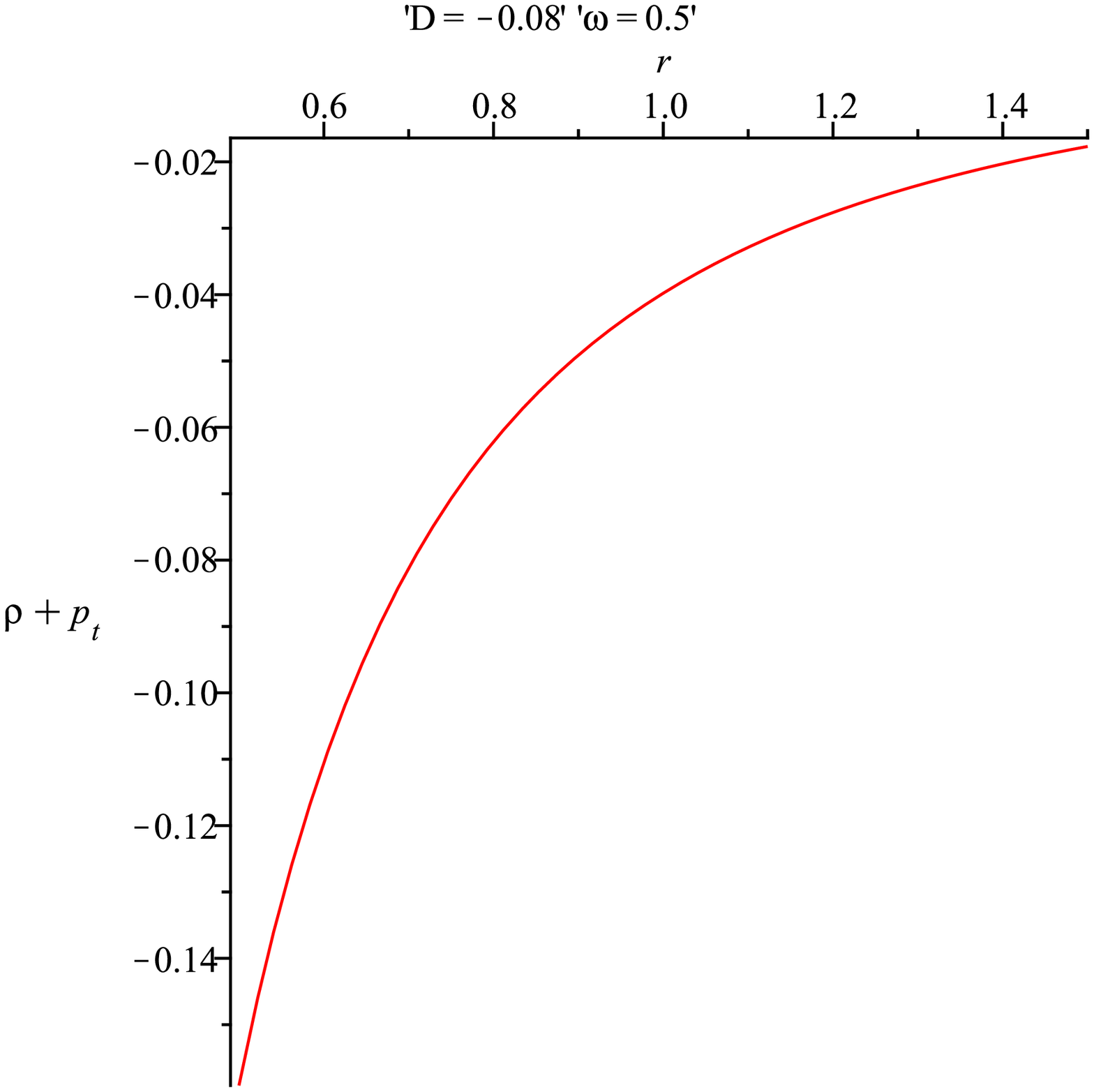}
\includegraphics[width=0.3\textwidth]{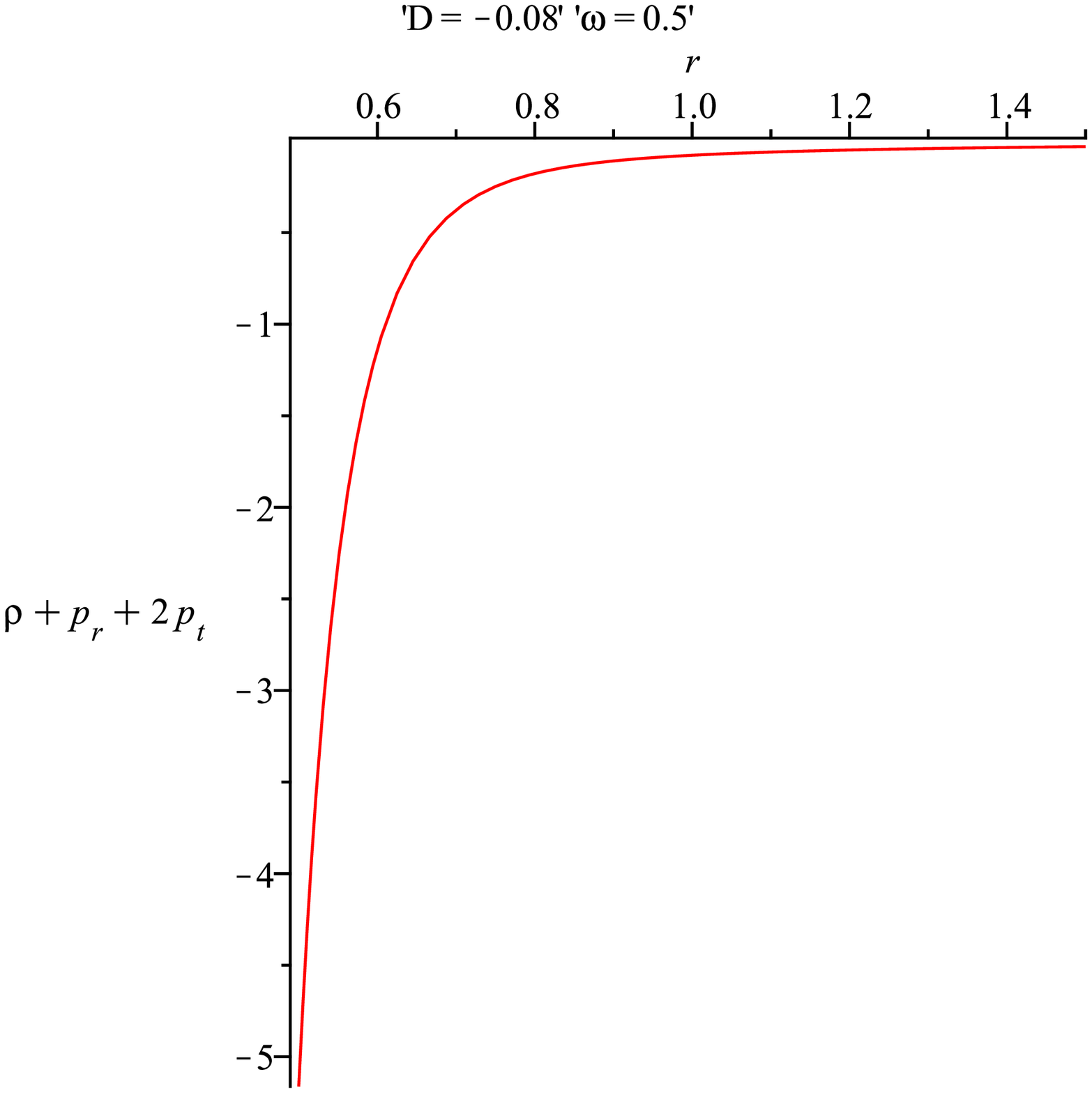}
\caption{Variation of $\rho + p_t$ (Top) and $\rho+p_r+2p_t$
(Bottom) are shown with respect to radial coordinate for
anisotropic case}
\end{figure}

\begin{equation}
T(r) = - \left[\frac{6(\omega+1)}{(\omega+3)}\right]r^{-2}
-6Dr^{-\left[\frac{3(\omega+1)}{\omega}\right]},\label{eq27}
\end{equation}

\begin{equation}
4 \pi \rho(r)=-\left[\frac{
3D}{2\omega}\right]r^{-\left[\frac{3(\omega+1)}{\omega}\right]}-\left[\frac{1}{\omega+3}\right]r^{-2},\label{eq28}
\end{equation}

\begin{equation}
4\pi
p_r=-\left[\frac{3D}{2}\right]r^{-\left[\frac{3(\omega+1)}{\omega}\right]}-\left[\frac{\omega}{\omega+3}\right]r^{-2},\label{eq29}
\end{equation}

\begin{equation}
4 \pi p_t=
\left[\frac{3D}{2\omega}\right]r^{-\left[\frac{3(\omega+1)}{\omega}\right]}-\left[\frac{(\omega+1)}{2(\omega+3)}\right]r^{-2},\label{eq30}
\end{equation}

\begin{equation}
\psi=  C \left[ \left(\frac{\omega+1}{\omega+3}\right)
+Dr^{-\left(\frac{\omega+3}{\omega}\right)}\right]^{\frac{1}{2}},\label{eq31}
\end{equation}

\begin{equation}
\frac{dp_t}{d\rho}=
\frac{-\frac{9D\left(\omega+1\right)}{2\omega^2}r^{-\left(\frac{\omega+3}{\omega}\right)}+\left(\frac{\omega+1}{\omega+3}\right)}
{\frac{9D\left(\omega+1\right)}{2\omega^2}r^{-\left(\frac{\omega+3}{\omega}\right)}+\left(\frac{2}{\omega+3}\right)},\label{eq32}
\end{equation}
where $D$ is an integration constant.

\begin{figure}[h]
\centering
\includegraphics[width=0.3\textwidth]{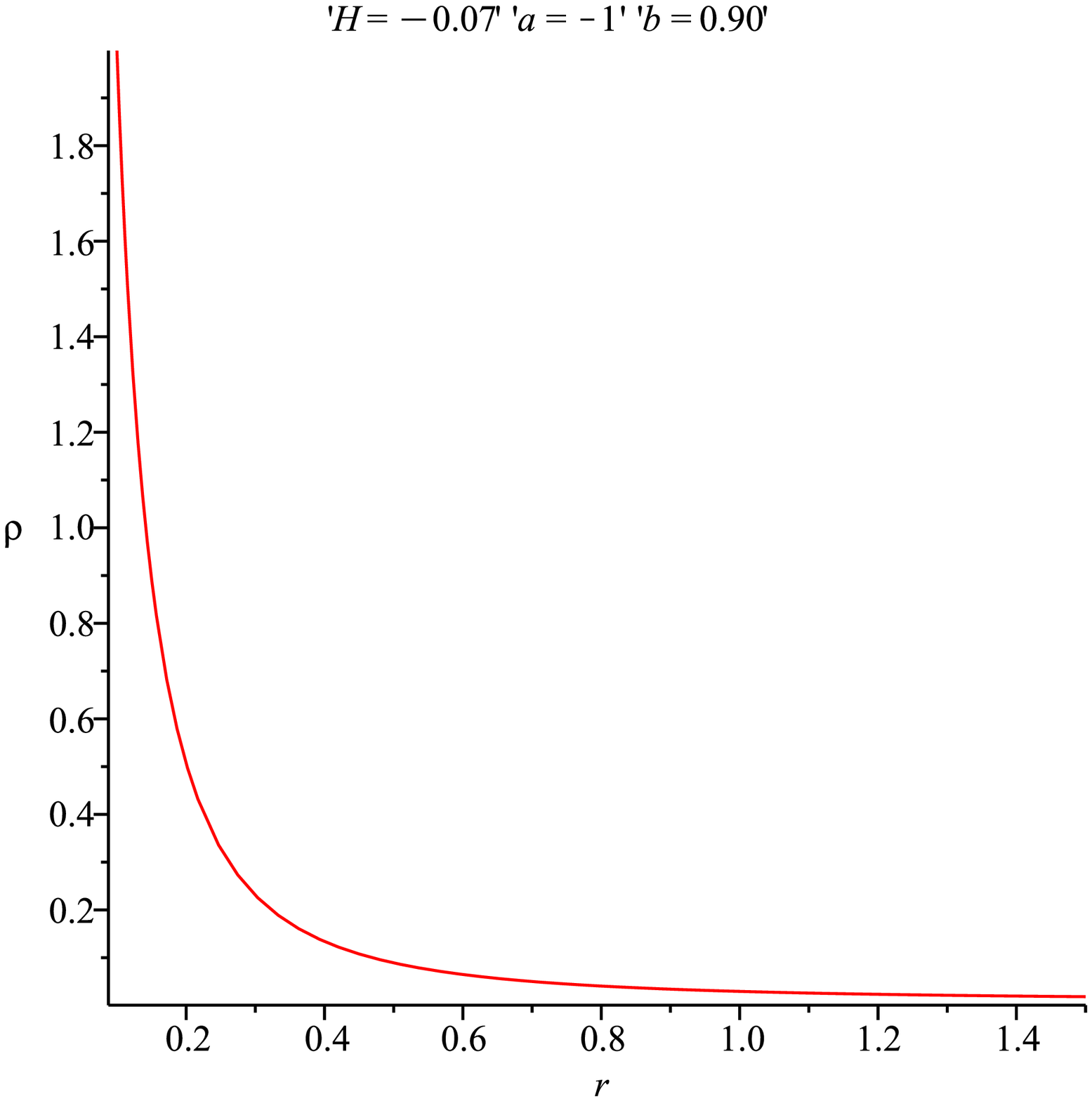}
\includegraphics[width=0.3\textwidth]{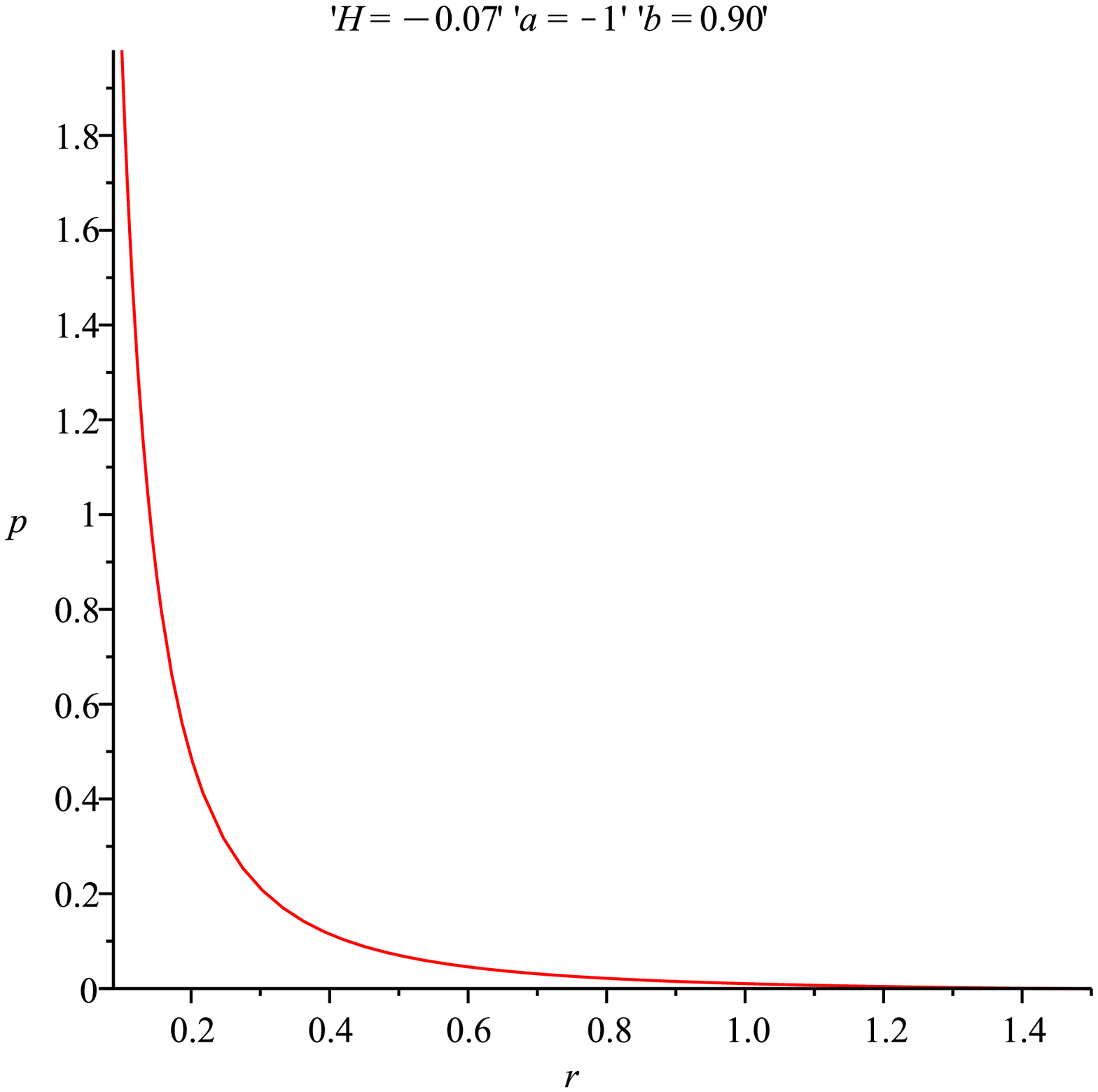}
\includegraphics[width=0.3\textwidth]{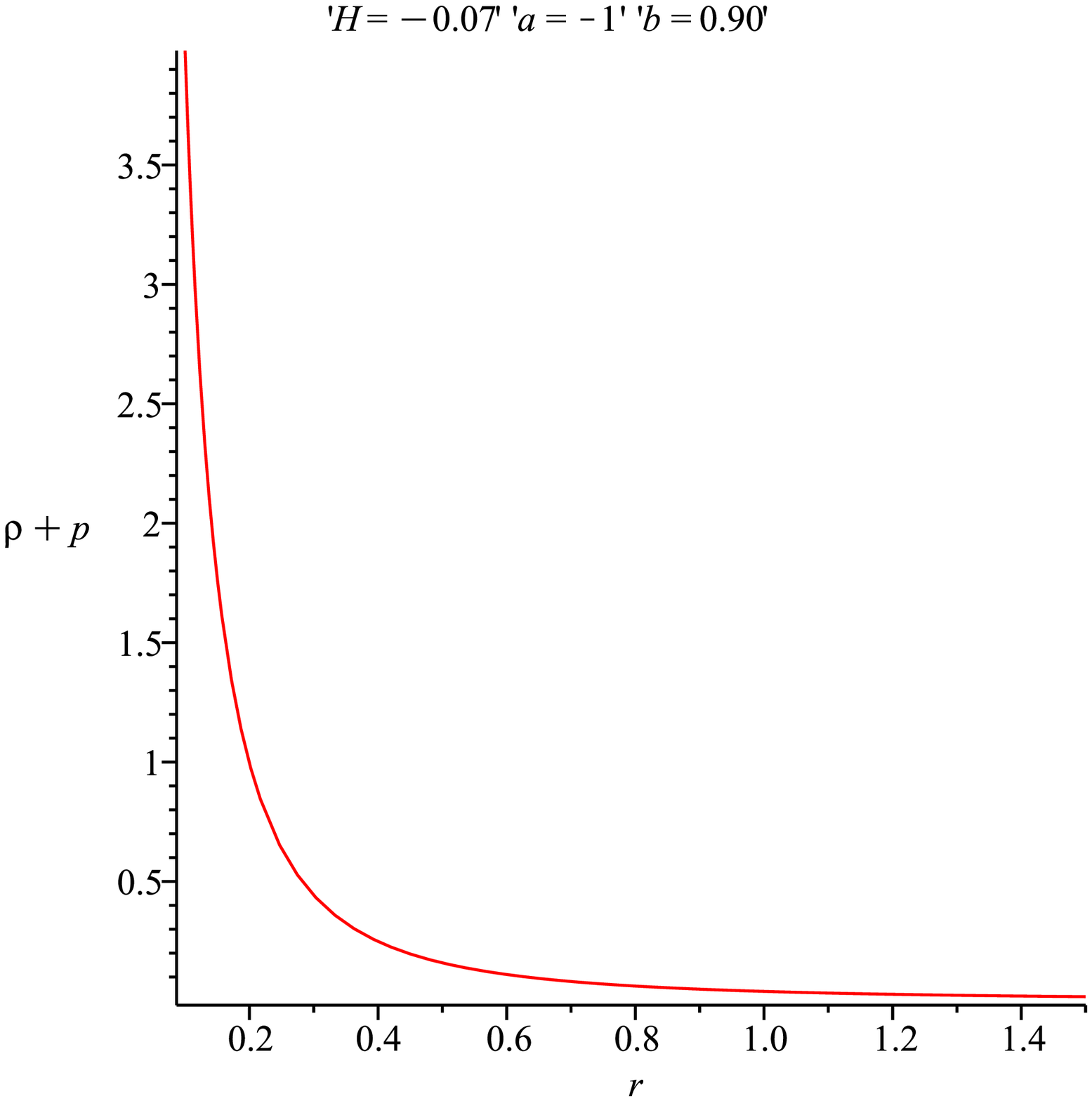}
\includegraphics[width=0.3\textwidth]{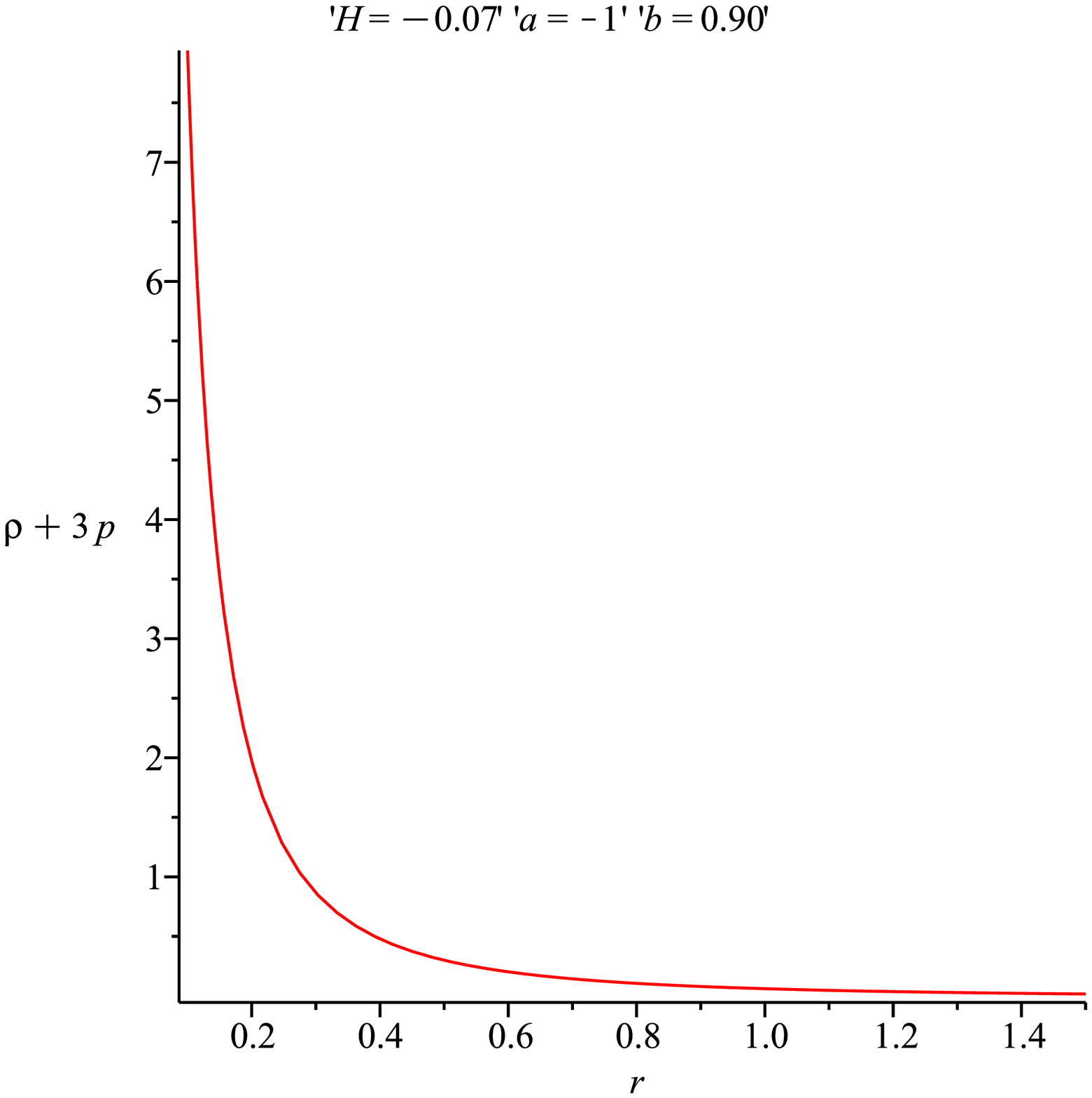}
\caption{Variation of $\rho$ (Top), $p$ (Upper Middle), $\rho+p$
(Lower Middle) and $\rho+3p$ (Bottom) are shown with respect to
radial coordinate for isotropic case.}
\end{figure}

\subsection{CASE II: $f(T)=aT+b$}

\subsubsection{$p_r = p_t = p$}

Now, using Eqs. (\ref{eq14}) - (\ref{eq17}) and (\ref{eq18b}), we
obtain the following solutions \citep{Herrera1984,Herrera1985c}:
\begin{equation}
e^{-\lambda(r)}=\left( \frac{1}{2}+H r^2\right),\label{eq33}
\end{equation}

\begin{equation}
T(r) = - \left[\frac{3}{r^2}+6H\right],\label{eq34}
\end{equation}

\begin{equation}
4 \pi \rho(r)=
-\frac{a}{4r^2}+\frac{3}{2}aH+\frac{b}{4},\label{eq35}
\end{equation}

\begin{equation}
4 \pi p =-\frac{a}{4r^2}-\frac{3}{2}aH-\frac{b}{4} ,\label{eq36}
\end{equation}

\begin{equation}
\psi=  C \left[ \frac{1}{2}+Hr^2\right]^{\frac{1}{2}},\label{eq37}
\end{equation}
where $a$ and $b$ are two purely constant quantities as mentioned
earlier and $H$ is an integration constant.

As before the sound velocity ${v_s}^2$ can be given by
\begin{equation}
\frac{dp}{d\rho}= 1.\label{eq38}
\end{equation}

\begin{figure}[h]
\centering
\includegraphics[width=0.3\textwidth]{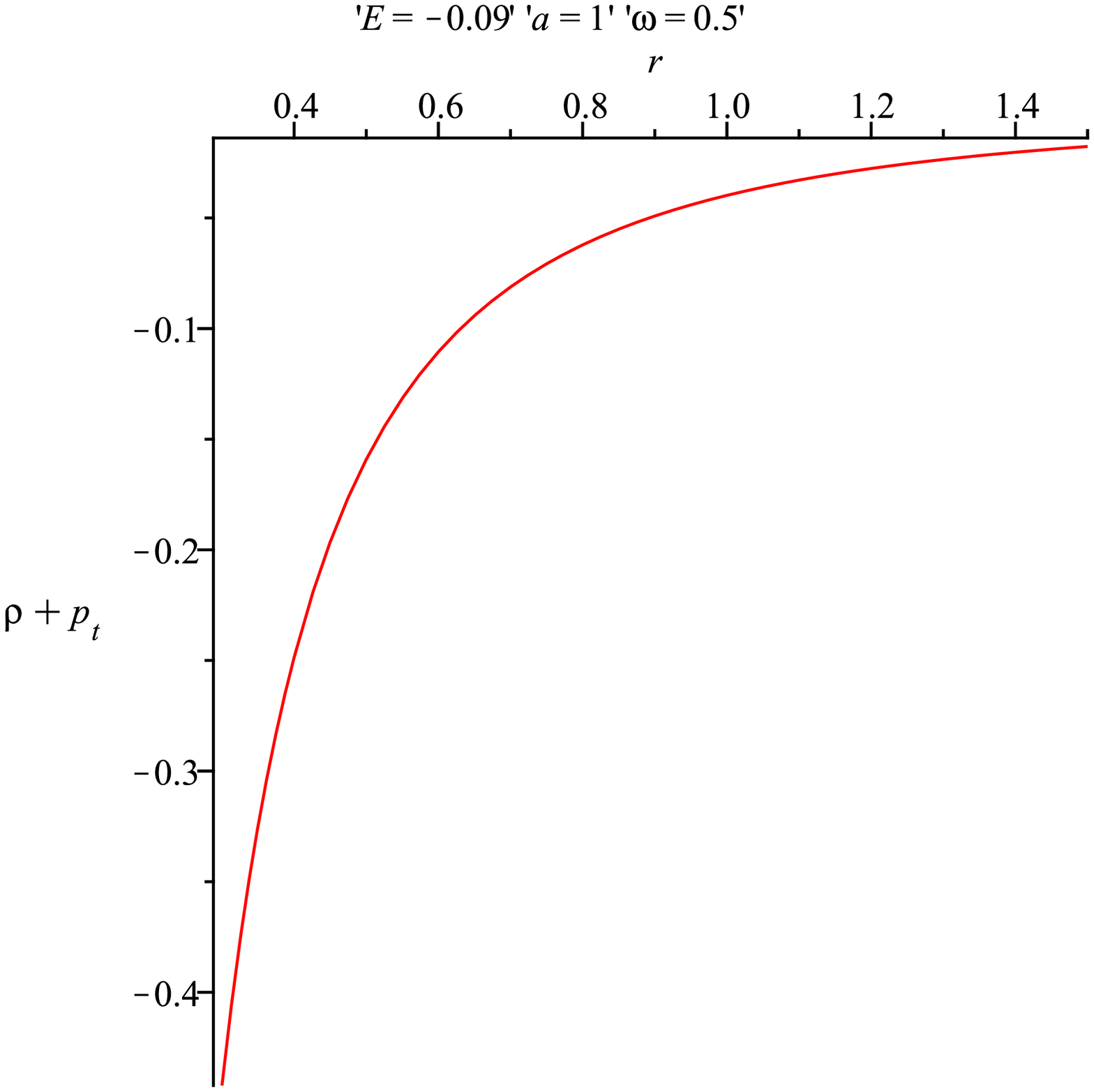}
\includegraphics[width=0.3\textwidth]{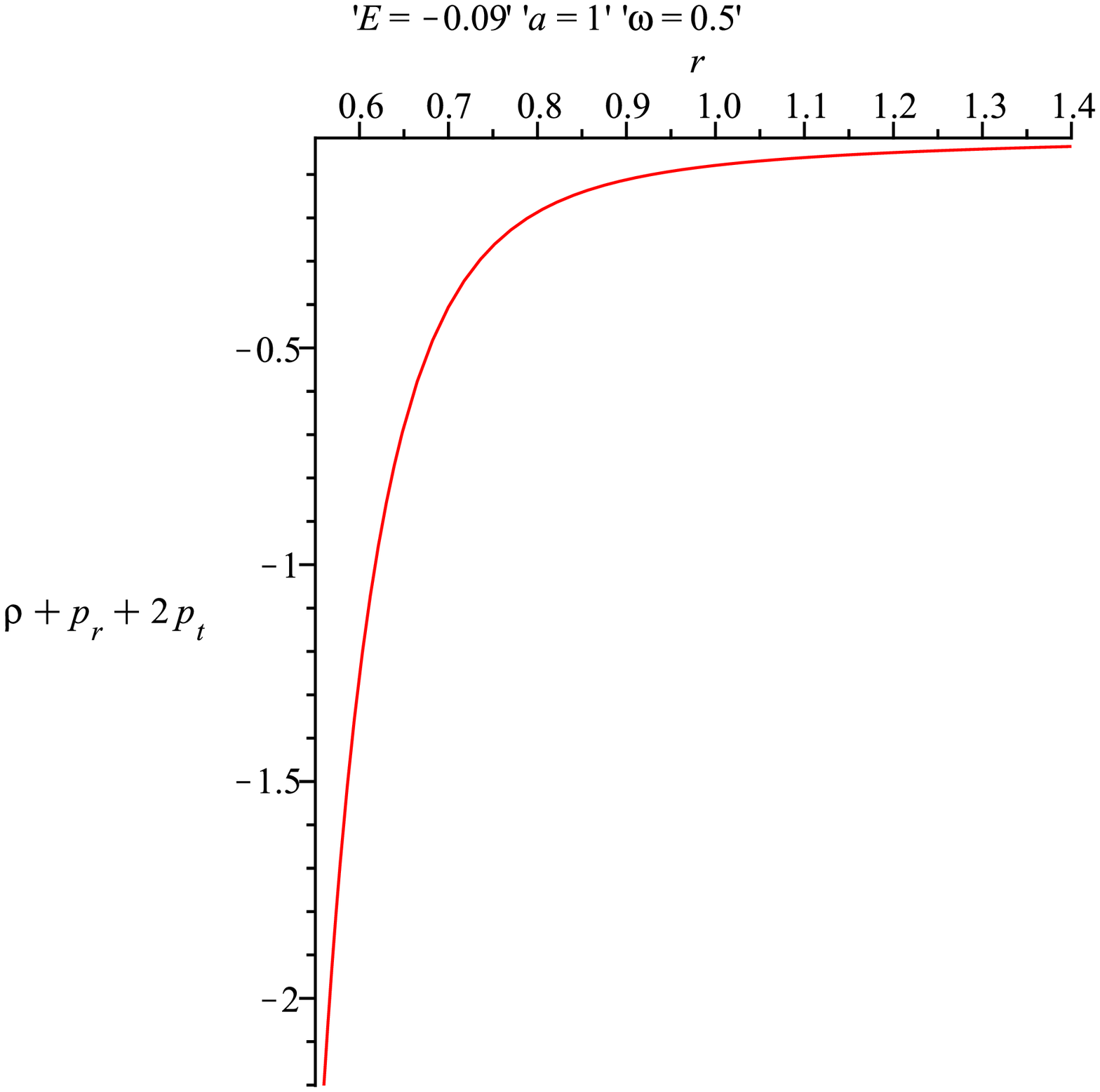}
\caption{Variation of $\rho + p_t$ (Top) and $\rho+p_r+2p_t$
(Bottom) are shown with respect to radial coordinate for
anisotropic case}
\end{figure}

\subsubsection{$p_r \neq p_t$}

Now, using Eqs. (\ref{eq14}) - (\ref{eq17}), (\ref{eq18b}) and
(\ref{eq19}), we obtain the following solutions
\citep{Herrera1984,Herrera1985c}:
\begin{equation}
e^{-\lambda(r)}=  \left[\frac{\omega+1}{\omega+3}\right]
+Er^{-\left[\frac{\omega+3}{\omega}\right]}-\frac{br^2}{6a},\label{eq39}
\end{equation}

\begin{equation}
T(r) = - \left[\frac{6(\omega+1)}{(\omega+3)}\right]r^{-2}
-6Er^{-\left[\frac{3(\omega+1)}{\omega}\right]}+\frac{b}{a},\label{eq40}
\end{equation}

\begin{equation}
4 \pi \rho(r)=-\left[\frac{
3aE}{2\omega}\right]r^{-\left[\frac{3(\omega+1)}{\omega}\right]}-\left[\frac{a}{\omega+3}\right]r^{-2},\label{eq41}
\end{equation}

\begin{equation}
4 \pi p_r=-\left[\frac{
3aE}{2}\right]r^{-\left[\frac{3(\omega+1)}{\omega}\right]}-\left[\frac{a\omega}{\omega+3}\right]r^{-2},\label{eq42}
\end{equation}

\begin{equation}
4 \pi p_t=\left[
\frac{3aE}{2\omega}\right]r^{-\left[\frac{3(\omega+1)}{\omega}\right]}
-\left[\frac{ a(\omega+1)}{2(\omega+3) }\right]r^{-2}
,\label{eq43}
\end{equation}

\begin{equation}
\psi=  C \left[ \left(\frac{\omega+1}{\omega+3}\right)
+Er^{-\left(\frac{\omega+3}{\omega}\right)}-\frac{br^2}{6a}\right]^{\frac{1}{2}},\label{eq44}
\end{equation}

\begin{equation}
\frac{dp_t}{d\rho}=
\frac{-\frac{9aE\left(\omega+1\right)}{2\omega^2}r^{-\left(\frac{\omega+3}{\omega}\right)}+\left(\frac{a(\omega+1)}{\omega+3}\right)}
{\frac{9aE\left(\omega+1\right)}{2\omega^2}r^{-\left(\frac{\omega+3}{\omega}\right)}+\left(\frac{2a}{\omega+3}\right)},\label{eq45}
\end{equation}
where $E$ is an integration constant.

\section{Physical features of the model}

\subsection{Energy conditions}

Now, let us check whether all the energy conditions are satisfied or
not for the present model under $f(T)$ gravity. For  this purpose,
we should consider the following inequalities:
\[
(i)~NEC: \rho+p_r\geq 0,~\rho+p_t\geq 0,
\]
\[
(ii)~WEC: \rho+p_r\geq 0,~\rho\geq 0,~\rho+p_t\geq 0,
\]
\[
(iii)~SEC: \rho+p_r\geq 0,~\rho+p_r+2p_t\geq 0.
\]

Among all the above CASES I and II of $f(T)$ gravity it is
revealed that only the solutions of sub-case with isotropic
condition ($p_r = p_t = p$) under CASE II i.e. $f(T)=aT+b$ (4.2.1)
are physically valid (see Fig. 3 for energy conditions). The other
cases are not physically interesting the energy conditions being
violated there (Figs. 1,~2 and 4).

\begin{figure}[h]
\centering
\includegraphics[width=0.3\textwidth]{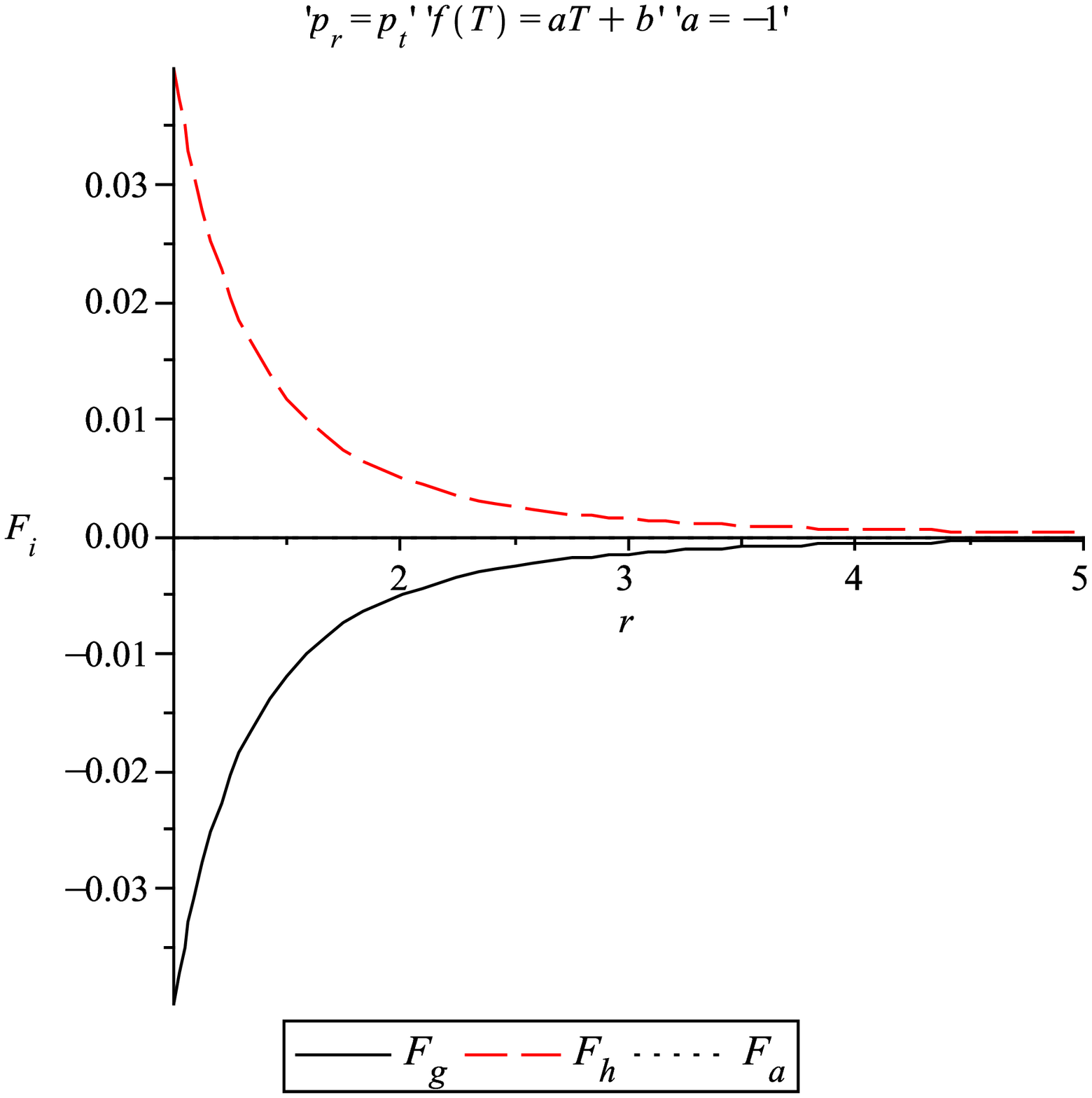}
\caption{The three different forces, viz. gravitational force
$(F_g)$, hydrostatic force $(F_h)$ and anisotropic force $(F_a)$
are plotted against $r$~(km) for the sub-case 4.2.1}
\end{figure}

\subsection{TOV equation}

The Generalized Tolman-Oppenheimer-Volkoff (TOV) equation can be
written in the form
\begin{equation}
-\frac{M_G(r)(\rho+p_r)}{r^2}e^{\frac{\lambda-\nu}{2}}-\frac{dp_r}{dr}+\frac{2}{r}(p_t-p_r)=0,\label{eq46}
\end{equation}
where $M_G(r)$ is the gravitational mass within the sphere of
radius $r$ and is given by
\begin{equation}
M_G(r)=\frac{1}{2}r^2e^{\frac{\nu-\lambda}{2}}\nu'.\label{eq47}
\end{equation}

Substituting Eq. (\ref{eq47}) into Eq. (\ref{eq46}), we obtain
\begin{equation}
-\frac{\nu'}{2}(\rho+p_r)-\frac{dp_r}{dr}+\frac{2}{r}(p_t-p_r)=0.\label{eq48}
\end{equation}

The above TOV equation describe the equilibrium of the stellar
configuration under the joint action of the different forces, viz.
gravitational force ($F_g$), hydrostatic force ($F_h$) and
anisotropic force ($F_a$) so that eventually as an equilibrium
condition one can write it in the following form:
\begin{equation}
F_g+F_h+F_a=0,\label{eq49}
\end{equation}
where
\[F_g=-\frac{\nu'}{2}(\rho+p_r),\]
\[F_h=-\frac{dp_r}{dr},\]
\[F_a=\frac{2}{r}(p_t-p_r).\]

In case of $p_r = p_t$ and $f(T)=aT+b$ the gravitational and
hydrostatic forces are given by
\begin{equation}
F_g = \frac{a}{8\pi r^3}
\end{equation}
and
\begin{equation}
F_h = - \frac{a}{8\pi r^3}.
\end{equation}

We have plotted the feature of TOV equation for the sub-case {\it
4.2.1} in Fig. 5 (we don't need to include TOV equation in the
sub-case {\it 4.1.1} as the model does not exist the energy
conditions being violated). In this sub-case of {\it 4.2.1} we
observe that static equilibrium has been attained by the forces
through balancing of the stresses between them. However, it can be
found out that anisotropic forces have no contribution to
equilibrium i.e. $F_a =0$ as $p_r =p_t$.

\begin{figure}[h]
\centering
\includegraphics[width=0.3\textwidth]{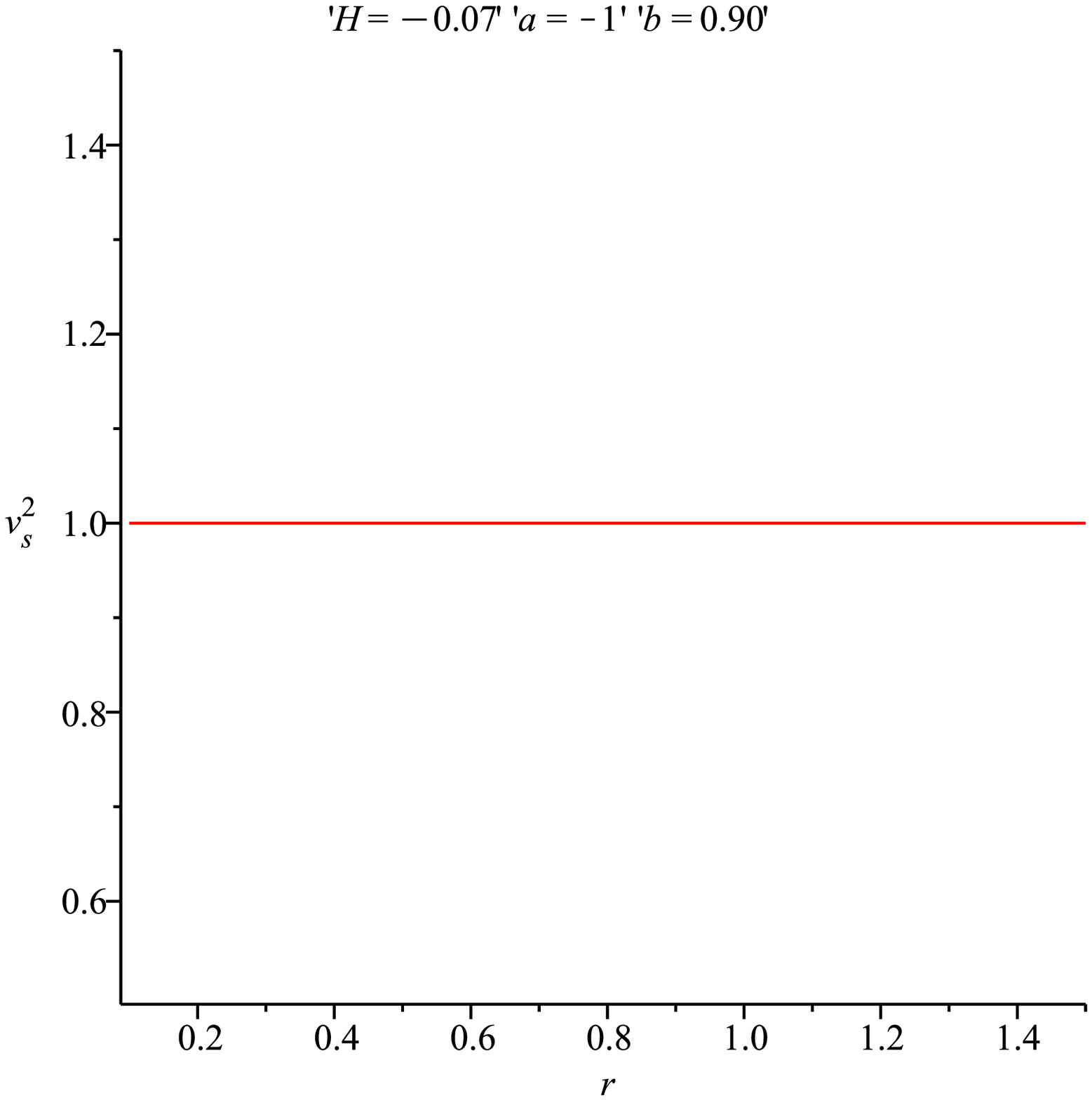}
\caption{Profile of sound velocity is shown which actually remains
constant.}
\end{figure}

\subsection{Stability issue}

We shall now turn our attention to the stability issue of model.
According to the cracking technique proposed by \cite{Herrera1992}
the squares of the sound speed should be within the limit $[0,1]$.
Fig. 6 satisfies Herrera's criterion i.e. ${v_s}^2 \geq 0$ within
the matter distribution and therefore our model maintains
stability.

\subsection{Nature of the star}

\begin{figure}[h]
\centering
\includegraphics[width=0.3\textwidth]{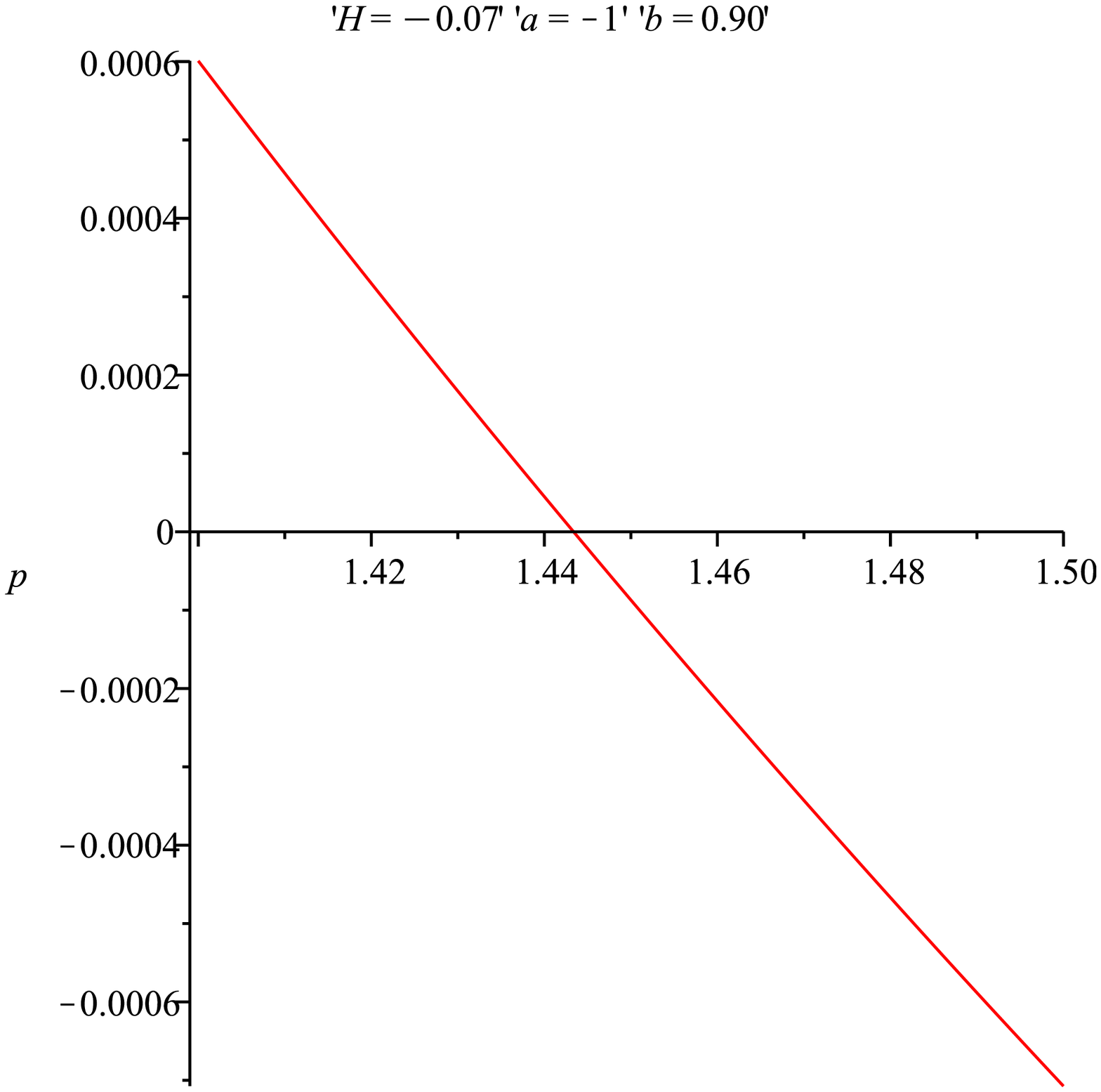}
\caption{Radius of the star is given where $p$ cuts $r$ axis.}
\end{figure}

We have drawn a plot to show radius of our stellar model. It can
be seen by the clear `cut' on X-axis which is turned out to be
1.445 km (Fig. 7). This is very small and indicates towards a
compact star with ultra-compactness. A tally of this value with
the already available data set immediately reveals that the star
is nothing but either a quark/strange star (see Table 1 of the
Ref. \cite{Bhar2015}) or a brown dwarf star of type F5 (see the
Link:
http://www.world-builders.org/lessons/less/les1/StarTables.html).

This value of $R=1.445$~km immediately provoke us to figure out
the surface density of the stellar system. As $r$ tends to zero,
$\rho$ goes to infinity and therefore, we are not in position to
comment on the central density. However, we can estimate the
surface density of the star by plugging $G$ and $c$ in the
expression of $\rho$ which eventually yields the numerical value
as $4.847 \times 10^{16}$~gm/cc. Therefore, this very high density
in the order $10^{16}$~gm/cc with a very small radius $R=1.445$~km
really indicates that the model under $f(T)$ gravity represents an
ultra-compact
star~\citep{Ruderman1972,Glendenning1997,Herjog2011}.

\section{Conclusions}

We have studied in detail the $f(T)$ gravity for the two specific cases
$f(T)= T$ and $f(T)=aT+b$. It has specially been observed that
among all the CASES I and II of $f(T)$ gravity only the solutions
of sub-case 4.2.1 with isotropic condition ($p_r = p_t = p$)
under CASE II, i.e. $f(T)=aT+b$ are physically valid. In general,
our observation is that under $f(T)$ gravity anisotropy doesn't exist
as well as all the energy conditions become jeopardized. Obviously,
in favor of this conclusion further studies are needed
to perform several other aspect of the $f(T)$ theory of gravity.

In connection to features and hence validity of the model
we have studied several physical aspects based on the solutions
and all these being very interesting advocate in favor of
physically acceptance of the model. We would like to summarize
all these results as follows:

{\bf (i) Energy conditions}

Our study reveals that only the solutions of sub-case 4.2.1
with isotropic condition under CASE II i.e. $f(T)=aT+b$
are physically valid as in the other cases energy
conditions are seen to be violated.

{\bf (ii) TOV equation}

The plot for the generalized TOV equation shows that static
equilibrium has been attained by the different forces, viz.
gravitational force ($F_g$), hydrostatic force ($F_h$) and
anisotropic force ($F_a$). However, it is also observed that
anisotropic forces have no role for equilibrium of the system.

{\bf (iii) Stability issue}

By employing the cracking concept of \cite{Herrera1992} we have
demonstrated through plot that the squares of the sound speed
remains within the limit $[0,1]$ and hence shown our model is a
stable one.

{\bf (iv) Nature of the star}

The model with very high density ($4.847 \times 10^{16}$~gm/cc)
and small radius ($R=1.445$~km) suggests that the present investigation
under $f(T)$ theory of gravity is a representative of an ultra-compact star.

We note an interesting work on the $f(T)$ gravity in connection to
\cite{Krori1975} metric parameters as done by \cite{Abbas2015}.
The present work therefore may be extended to that line of
thinking in a future project.

\section*{Acknowledgments} FR and SR are thankful to the
Inter-University Centre for Astronomy and Astrophysics (IUCAA),
India for providing Visiting Associateship under which a part of
this work was carried out.


\begin{thebibliography}{0}

\bibitem[Abbas et al. (2015)]{Abbas2015} Abbas, G., Kanwal, A., Zubair, M.: arXiv: 1501.05829 [physics.gen-ph] (2015)

\bibitem[Aftergood and DeBenedictis (2014)]{Aftergood2014} Aftergood, J., DeBenedictis,
A.: Phys. Rev. D {\bf 90} 124006 (2014)

\bibitem[Aldrovandi and Pereira (2013)]{Aldrovandi2013} Aldrovandi, R., Pereira, J.G.: Teleparallel Gravity: An
Introduction, (Springer, Dordrecht - Heidelberg - New York -
London, 2013)

\bibitem[Bamba et al. (2011)]{Bamba2011} K. Bamba, C.-Q. Geng, C.C. Lee, L.-W. Luo: JCAP {\bf 1101}, 021 (2011)

\bibitem[Bengochea (2011)]{Bengochea2011} Bengochea, G.R.: Phys. Lett. B {\bf 695}, 405 (2011)

\bibitem[Bengochea and Ferraro (2009)]{Bengochea2009} Bengochea, G.R., Ferraro, R.: Phys. Rev. D {\bf 79}, 124019 (2009)

\bibitem[Bhar (2014)]{Bhar2014} Bhar, P.: Astrophys. Space Sci. \textbf{354}, 457 (2014)

\bibitem[Bhar et al. (2015)]{Bhar2015} Bhar, P., Rahaman, F., Ray, S., Chatterjee, V.: Eur. Phys. J. C {\bf 75}, 190 (2015)

\bibitem[B\"{o}hmer et al. (2007)]{Harko1} B\"{o}hmer, C.G., Harko, T., Lobo, F.S.N.: Phys. Rev. D {\bf 76}, 084014 (2007)

\bibitem[B\"{o}hmer et al. (2008)]{Harko2} B\"{o}hmer, C.G., Harko, T., Lobo, F.S.N.: Class. Quantum Gravit. {\bf 25}, 075016 (2008)

\bibitem[B{\"o}hmer et al. (2011)]{Boehmer2011} B{\"o}hmer, C.G., Mussa, A., Tamanini, N.: Class. Quantum Gravit. {\bf 28}, 245020 (2011)

\bibitem[Chen et al. (2011)]{Chen2011} Chen, S.-H., Dent, J.B., Dutta, S., Saridakis, E.N.: Phys. Rev. D {\bf 83}, 023508 (2011)

\bibitem[Daouda et al. (2011)]{Daouda2011} Daouda, M.H., Rodrigues, M.E., Houndjo, M.J.S.: Eur. Phys. J. C {\bf 71}, 1817 (2011)

\bibitem[De Andrade et al. (2000)]{Andrade2000} De Andrade, V.C., Guillen, L.C.T., Pereira, J.G.: Proc. IX Marcel Grossman
(Rome, Italy) (2000)

\bibitem[De Felice and Tsujikawa (2010)]{Felice2010} De Felice, A., Tsujikawa, S.: Living Rev. Rel. {\bf 13}, 3 (2010)

\bibitem[Deliduman and Yapiskan (2011)]{Deliduman2011} Deliduman, C., Yapiskan, B.: arXiv:1103.2225 [gr-qc] (2011)

\bibitem[Dent et al. (2011)]{Dent2011} Dent, J.B., Dutta, S., Saridakis, E.N.: JCAP {\bf 1101}, 009 (2011)

\bibitem[Durrer and Maartens (2010)]{Durrer2010} Durrer, R., Maartens, R.: Dark energy: Observational and theoretical approaches,
Ed. P. Ruiz-Lapuente (Cambridge UP, 2010)

\bibitem[Ferraro and Fiorini (2011)]{Ferraro2011} Ferraro, R., Fiorini, F.: Phys. Lett. B {\bf 702}, 75 (2011)

\bibitem[Glendenning (1997)]{Glendenning1997} Glendenning, N.K.: Compact Stars: Nuclear Physics,
Particle Physics and General Relativity (Springer-Verlag, New
York, p. 70, 1997)

\bibitem[Herjog and Roepke (2011)]{Herjog2011} Herjog, M., Roepke, F.K.: arXiv: 1109.0539 [astro-ph.HE]
(2011)

\bibitem[Herrera (1992)]{Herrera1992} Herrera, L.: Phys. Lett. A {\bf 165}, 206 (1992)

\bibitem[Herrera et al. (1984)]{Herrera1984} Herrera, L., J. Jimenez, J., L. Leal, L., Ponce de Leon, J., Esculpi, M.,
Galina, V.: J. Math. Phys. {\bf 25}, 3274 (1984)

\bibitem[Herrera and Ponce de Leon (1985a)]{Herrera1985a} Herrera, L., Ponce de Leon, J.: J. Math. Phys. {\bf 26}, 778 (1985a)

\bibitem[Herrera and Ponce de Leon (1985b)]{Herrera1985b} Herrera, L., Ponce de Leon, J.: J. Math. Phys. {\bf 26}, 2018 (1985b)

\bibitem[Herrera and Ponce de Leon (1985c)]{Herrera1985c} Herrera, L., Ponce de Leon, J.: J. Math. Phys. {\bf 26}, 2302 (1985c)

\bibitem[Krori and Barua (1975)]{Krori1975} Krori, K.D., Barua, J.: J. Phys. A.: Math. Gen.
{\bf 8}, 508 (1975)

\bibitem[Li et al. (2011)]{Li2011} Li, B., Sotiriou, T.P., Barrow, J.D.: Phys. Rev. D {\bf 83}, 064035 (2011)

\bibitem[Linder (2010)]{Linder2010} Linder, E.V.: Phys. Rev. D {\bf 81}, 127301 (2010)

\bibitem[Rahaman et al. (2010a)]{Rahaman2010a} Rahaman, F., Jamil, M., Sharma, R., Chakraborty, K.: Astrophys. Space Sci. {\bf 330}, 249 (2010)

\bibitem[Rahaman et al. (2010b)]{Rahaman2010b} Rahaman, F., Jamil, M., Kalam, M., Chakraborty, K., Ghosh, A.: Astrophys. Space Sci. {\bf 137}, 325 (2010)

\bibitem[Rahaman et al. (2014)]{Rahaman2014} Rahaman, F., et al.: Int. J. Mod. Phys. D {\bf 23}, 1450042
 (2014)

\bibitem[Rahaman et al. (2015a)]{Rahaman2015a} Rahaman, F., Karmakar, S., Karar, I., Ray, S.: Phys. Lett. B {\bf 746},
73 (2015)

\bibitem[Rahaman et al. (2015b)]{Rahaman2015b} Rahaman, F., Ray, S., Khadekar, G.S., Kuhfittig, P.K.F., Karar, I.: Int. J. Theor. Phys. {\bf 54}, 699 (2015)

\bibitem[Rahaman et al. (2015c)]{Rahaman2015c} Rahaman, F., Pradhan, A., Ahmed, N., Ray, S., Saha, B., Rahaman, M.:
Int. J. Mod. Phys. D {\bf 24}, 1550049 (2015)

\bibitem[Ray et al. (2008)]{Ray2008} Ray, S., Usmani, A.A., Rahaman, F., Kalam, M., Chakraborty, K.: Ind. J. Phys. {\bf 82}, 1191 (2008)

\bibitem[Ruderman (1972)]{Ruderman1972} Ruderman, R.: Rev. Astr. Astrophys. {\bf 10}, 427
(1972)

\bibitem[Sotiriou and Faraoni (2010)]{Sotiriou2010} Sotiriou, T.P., Faraoni, V.: Rev. Mod. Phys. {\bf 82}, 451 (2010)

\bibitem[Tamanini (2012)]{Tamanini2012a} Tamanini, N.: Proc. 13th Marcell Grossman Meeting,
University College London (2012)

\bibitem[Tamanini and B{\"o}hmer (2012)]{Tamanini2012b} Tamanini, N., B{\"o}hmer, C.G.: Phys. Rev. D {\bf 86}, 044009 (2012)

\bibitem[Tsyba et al. (2011)]{Tsyba2011} Tsyba, P.Y., Kulnazarov, I.I., Yerzhanov, K.K., Myrzakulov, R.: Int. J. Theor. Phys. {\bf 50}, 1876 (2011)

\bibitem[Usmani et al. (2011)]{Usmani2011} Usmani, A.A., Rahaman, F., Ray, S., Nandi, K.K., Kuhfittig, P.K.F., Rakib, Sk.A., Hasan, Z.: Phys. Lett. B {\bf
701}, 388 (2011)

\bibitem[Wang (2011)]{Wang2011} Wang, T.: Phys. Rev. D {\bf 84}, 024042 (2011)

\bibitem[Wu and Yu (2010a)]{Wu2010a} Wu, P., Yu, H.W.: Phys. Lett. B {\bf 692}, 176 (2010)

\bibitem[Wu and Yu (2010b)]{Wu2010b} Wu, P., Yu, H.W.: Phys. Lett. B {\bf 693}, 415 (2010)

\bibitem[Wu and Yu (2011)]{Wu2011} Wu, P., Yu, H.W.: Eur. Phys. J. C {\bf 71}, 1552 (2011)

\bibitem[Yang (2011)]{Yang2011} Yang, R.-J.: Europhys. Lett. {\bf 93}, 60001 (2011)

\bibitem[Zhang et al. (2011)]{Zhang2011} Zhang, Y., Li, H., Gong, Y., Zhu, Z.-H.: JCAP {\bf 1107}, 015 (2011)

\end{thebibliography}
\end{document}